\def\ra{\rangle}
\def\la{\langle}
\def\be{\begin{equation}}
\def\ee{\end{equation}}
\def\ba{\begin{array}}
\def\ea{\end{array}}

\documentclass[prl,showpacs,twocolumn,amsmath]{revtex4}
\usepackage{epsfig,graphicx}
\usepackage{amsmath}
\input amssym.def

\begin{document}

\title{Local Unitary Equivalence of Multi-qubit Mixed quantum States}
\author{Ming Li$^{1,2}$}
\author{Tinggui Zhang$^{2}$}
\author{Shao-Ming Fei$^{2,3}$}
\author{Xianqing Li-Jost$^{2}$}
\author{Naihuan Jing$^{2,4,5}$}

\affiliation{$^1$School of Science, China University of
Petroleum, 266580 Qingdao, China\\
$^2$Max-Planck-Institute for Mathematics in the Sciences, 04103
Leipzig, Germany\\
$^3$Department of Mathematics, Capital Normal University, 100037
Beijing, China\\
$^4$Department of Mathematics, North Carolina State University,
Raleigh, NC27695, USA\\
$^5$School of Science, South China University of Technology,
Guangzhou 510640, China}

\begin{abstract}
We present computable criterion for completely classifying
multi-qubit quantum states under local unitary operations. The
criterion can be used to detect whether two quantum states in
multi-qubit systems are local unitary equivalent or not. Once
obtaining the positive answer, we are further able to compute the
corresponding unitary operators precisely. Since the scheme is based
on the mean values of some quantum mechanical observables, it
supplies an experimental way to judge the local equivalence of
quantum states.
\end{abstract}

\pacs{03.67.-a, 02.20.Hj, 03.65.-w}
\maketitle

{\it{Introduction:}} Quantum entangled states have become the most
important physical resource  in the rapidly developing field of
quantum information science \cite{nielsen}. In particular,
multipartite quantum entanglement plays key roles in many quantum
information processing like one-way quantum computing, quantum error
correction and quantum secret sharing \cite{1,2,3,4}. However, the
nonlocal properties of multipartite mixed states are much more
difficult to clarify than bipartite or pure states. Since two
quantum states will be of the same power in implementing quantum
information processing if they can be transferred to each other by
local unitary (LU) transformations, and many crucial characters such
as the degree of entanglement \cite{eof1,eof2}, the maximal
violations of Bell inequalities \cite{bell1,bell2,bell3,bell4}, and
the teleportation fidelity \cite{tel1,tel2} remain invariant under
the LU transformations, it has been an important problem to give a
complete and operational classification of quantum states under LU
transformations.

In principle, the LU equivalence problem can be characterized by the
complete set of invariants under local unitary transformations.
The author in \cite{makhlin} presented a
complete set of 18 polynomial invariants for the locally unitary
equivalence of two-qubit mixed states. Nice results have been
obtained for three qubits states \cite{linden, Linden}, some generic
mixed states \cite{SFG, SFW, SFY}, tripartite pure and mixed states
\cite{SCFW}. In \cite{mqubit}, Kraus presented a way to
determine the local unitary equivalence of $n$-qubit pure states,
which is extended to general multipartite pure states of arbitrary dimensions \cite{bliu}.

Mixed states are more general physically. For bipartite mixed
quantum systems, Zhou et.al \cite{zhou} have solved the local
unitary equivalence problem by presenting a complete set of
invariants such that two density matrices are locally equivalent if
and only if all these invariants have equal values in these density
matrices, which is operational for non-degenerate states. In
\cite{zhang}, we have investigated the LU equivalence problem in
terms of matrix realignment and partial transposition. A necessary
and sufficient criterion for the local unitary equivalence of
multipartite states, together with explicit forms of the local
unitary operators have been presented. The criterion is shown to be
operational for states having eigenvalues with multiplicity of no
more than 2. However, as the two criteria above depend on the
eigenvectors of density matrices, for degenerate states(especially
for those with degeneracy larger than 2), they are less operational,
similar to the multipartite pure state case when the reduced density
matrices or the co-tensor matrices are degenerate.

In this paper, we investigate LU equivalence criterion and
classification for general multipartite mixed quantum states.
Alternatively, we deal with problem in terms of the generalized
Bloch representations of density matrices. Hence the results work no
matter the density matrices are degenerate or not. In particular, we
present a both sufficient and necessary LU criterion for multi-qubit
systems. To show that the criterion is operational, we give a
detailed process in classifying multi-qubit mixed states under LU
transformations. Moreover, our criterion can not only verify the LU
equivalence of two mixed states, but also gives rise to the
corresponding LU operators that transform one state to another LU
equivalent one.\\

{\it{ Local unitary equivalence for multipartite states:}}~ We first
consider the local unitary equivalence problem for general
$H_1\otimes H_2\otimes\cdots\otimes H_N$ quantum systems with
$dim\,H_i=d_i$, $i=1,2,\cdots,N$. A multipartite state $\rho\in
H_1\otimes H_2\otimes\cdots\otimes H_N$ can be generally expressed
in terms of the $SU(n)$ generators $\lambda_{\alpha_{k}}$
\cite{hassan},

\begin{widetext}
\be\label{OS} \ba{rcl}
\rho&=&\displaystyle\frac{1}{\Pi_{i}^{N}d_{i}}\otimes_{j}^{N}I_{d_{j}}
+\sum\limits_{M=1}^N\sum\limits_{\{\mu_{1}\mu_{2}\cdots\mu_{M}\}}\sum\limits_{\alpha_{1}\alpha_{2}\cdots\alpha_{M}}
{\mathcal{T}}_{\alpha_{1}\alpha_{2}\cdots\alpha_{M}}^{\{\mu_{1}\mu_{2}\cdots\mu_{M}\}}\lambda_{\alpha_{1}}
^{\{\mu_{1}\}}\lambda_{\alpha_{2}}^{\{\mu_{2}\}}\cdots\lambda_{\alpha_{M}}^{\{\mu_{M}\}}
\ea \ee
\end{widetext}
where $I_{d_{j}}$ denotes ${d_{j}}\times {d_{j}}$ identity matrices,
$\lambda_{\alpha_{k}}^{\{\mu_{k}\}}=I_{d_{1}}\otimes
I_{d_{2}}\otimes\cdots\otimes
I_{d_{\mu_{k}-1}}\otimes\lambda_{\alpha_{k}}\otimes
I_{d_{\mu_{k}+1}}\otimes\cdots\otimes I_{d_{N}}$, with
$\lambda_{\alpha_{k}}$ appearing at the $\mu_k$th position, and

\begin{eqnarray*}
{\mathcal{T}}_{\alpha_{1}\alpha_{2}\cdots\alpha_{M}}
^{\{\mu_{1}\mu_{2}\cdots\mu_{M}\}}=\frac{\prod_{i=1}^{M}
d_{\mu_{i}}}{2^{M}}{\rm Tr}[\rho\lambda_{\alpha_{1}}
^{\{\mu_{1}\}}\lambda_{\alpha_{2}}^{\{\mu_{2}\}}\cdots\lambda_{\alpha_{M}}^{\{\mu_{M}\}}],
\end{eqnarray*}
which can be viewed as the entries of the tensors
${\mathcal{T}}^{\{\mu_{1}\mu_{2}\cdots\mu_{M}\}}$.

Before showing how to detect and classify quantum states under local
unitary equivalence, we give a short review of high order singular
value decomposition developed in \cite{siam}. For any tensor
${\mathcal{T}}$ with order $d_1\times d_2\times\cdots\times d_N$,
there exists a core tensor $\Sigma$ such that \be
{\mathcal{T}}=U^{(1)}\otimes U^{(2)}\otimes\cdots\otimes
U^{(N)}\Sigma, \ee where $\Sigma$ forms a same order tensor with
${\mathcal{T}}$. To calculate the core tensor $\Sigma$, one first
expresses ${\mathcal{T}}$ in matrix unfolding form
${\mathcal{T}}_n$. Then one derives the singular value decomposition
of the matrix ${\mathcal{T}}_n=U^{(n)}\Lambda^{(n)} V^{(n)}$. The
core tensor is then constructed by \be
\Sigma=\otimes_{n=1}^{N}U^{(n)\dag}{\mathcal{T}}. \ee
 Let $\rho$ and
$\rho'$ be two mixed states in $H_1\otimes H_2\otimes\cdots\otimes
H_N$. They are local unitary equivalent if
\be\label{dd}\rho'=(U_1\otimes...\otimes U_N)\rho
(U_1\otimes...\otimes U_N)^\dag\ee for some unitary operators $U_i$,
$i=1,2,...,N$, where $\dag$ denotes transpose and conjugate.\\

{\bf{Proposition:}} If $\rho$ and $\rho'$ are local unitary
equivalent, then their corresponding tensors
${\mathcal{T}}^{\{\mu_{1}\mu_{2}\cdots\mu_{M}\}}$ and
$({\mathcal{T}}')^{\{\mu_{1}\mu_{2}\cdots\mu_{M}\}}$ must have the
same core tensor up to the local symmetry
$\otimes_{n=1}^{M}P^{\mu_n}_{\{\mu_{1}\mu_{2}\cdots\mu_{M}\}}$ for
any $\{\mu_{1}\mu_{2}\cdots\mu_{M}\}\subset\{1,2,\cdots,N\}$, where
$P^{\mu_n}_{\{\mu_{1}\mu_{2}\cdots\mu_{M}\}}$ are block-diagonal
matrices with each block being orthogonal matrix of size equal to
the degeneracies of the singular values corresponding to the
${\mu_n}$ order unfolding of the tensor
${\mathcal{T}}^{\{\mu_{1}\mu_{2}\cdots\mu_{M}\}}$.\\

{\bf{Proof:}} Assume $\rho$ and $\rho'$ are related by (\ref{dd}).
Note that for any given unitary operator $U$, $U\lambda_{i}U^{\dag}$
is a traceless Hermitian operator which can be expanded according to
the $SU(d)$ generators,
\begin{eqnarray}\label{rr}
U\lambda_{i}U^{\dag}=\sum\limits_{j=1}^{d^{2}-1}\frac{1}{2}Tr\{U\lambda_{i}U^{\dag}\lambda_{j}\}\lambda_{j}\equiv
\sum\limits_{j=1}^{d^{2}-1}O_{ij}\lambda_{j}.
\end{eqnarray}
The entries $O_{ij}$ defines a real $(d^{2}-1)\times(d^{2}-1)$
matrix $O$. From the completeness relation of $SU(d)$ generators
$$
\sum\limits_{j=1}^{d^{2}-1}(\lambda_{j})_{ki}(\lambda_{j})_{mn}=2\delta_{im}\delta_{kn}
-\frac{2}{d}\delta_{ki}\delta_{mn},
$$
one can show that $O$ is an orthogonal matrix with determinant $+1$,
i.e. $O$ belongs to special orthogonal matrix group $SO(d)$.
According to (\ref{OS}) and (\ref{rr}), we get that there must exist
orthogonal operators $O^1$, $O^2,\cdots,O^N$ such that the following
equations hold,
\begin{widetext}
$$\ba{rcl}
\rho'&=&\frac{1}{\Pi_{i}^{N}d_{i}}\otimes_{j}^{N}I_{d_{j}}
+\sum\limits_{M=1}^N\sum\limits_{\{\mu_{1}\mu_{2}\cdots\mu_{M}\}}\sum\limits_{\alpha_{1}\alpha_{2}\cdots\alpha_{M}}
{\mathcal{T}}_{\alpha_{1}\alpha_{2}\cdots\alpha_{M}}^{\{\mu_{1}\mu_{2}\cdots\mu_{M}\}}\otimes_{n=1}^MU_{\mu_n}
\lambda_{\alpha_{1}}
^{\{\mu_{1}\}}\lambda_{\alpha_{2}}^{\{\mu_{2}\}}\cdots\lambda_{\alpha_{M}}^{\{\mu_{M}\}}\otimes_{n=1}^MU_{\mu_n}^{\dag}\\[4mm]
&=&\displaystyle\frac{1}{\Pi_{i}^{N}d_{i}}\otimes_{j}^{N}I_{d_{j}}
+\sum\limits_{M=1}^N\sum\limits_{\{\mu_{1}\mu_{2}\cdots\mu_{M}\}}\sum\limits_{\alpha_{1}\cdots\alpha_{M}\beta_{1}\cdots\beta_{M}}
{\mathcal{T}}_{\alpha_{1}\alpha_{2}\cdots\alpha_{M}}^{\{\mu_{1}\mu_{2}\cdots\mu_{M}\}}O^{\mu_1}_{\alpha_1\beta_1}O^{\mu_2}_{\alpha_2\beta_2}\cdots
O^{\mu_M}_{\alpha_M\beta_M}\lambda_{\beta_{1}}
^{\{\mu_{1}\}}\lambda_{\beta_{2}}^{\{\mu_{2}\}}\cdots\lambda_{\beta_{M}}^{\{\mu_{M}\}}\\[4mm]
\ea$$
\end{widetext}
Since $\rho'$ can also be written in the form of (\ref{OS}), we
obtain that
$$
{\mathcal{T'}}_{\beta_{1}\cdots\beta_{M}}
^{\{\mu_{1}\cdots\mu_{M}\}}=\sum_{\alpha_1\cdots\alpha_M}O^{\mu_1}_{\alpha_1\beta_1}\cdots
O^{\mu_M}_{\alpha_M\beta_M}{\mathcal{T}}_{\alpha_{1}\cdots\alpha_{M}}
^{\{\mu_{1}\cdots\mu_{M}\}}.
$$
There must exist orthogonal operators $O^1, O^2,\cdots,O^N$ such
that for any $1<M\leq N$,
\be\label{oo}{\mathcal{T'}}^{\{\mu_{1}\mu_{2}\cdots\mu_{M}\}}=\otimes_{n=1}^M
O^{\mu_n}{\mathcal{T}}^{\{\mu_{1}\mu_{2}\cdots\mu_{M}\}}.\ee By
virtue of Proposition 2 in \cite{bliu} and the analysis therein, one
has that the core tensor of the high order tensors
${\mathcal{T'}}^{\{\mu_{1}\mu_{2}\cdots\mu_{M}\}}$ and
${\mathcal{T}}^{\{\mu_{1}\mu_{2}\cdots\mu_{M}\}}$ must be the same
up to the local symmetry
$\otimes_{n=1}^{M}P^{\mu_n}_{\{\mu_{1}\mu_{2}\cdots\mu_{M}\}}$.
\hfill \rule{1ex}{1ex}\\

Let
\be\label{st}{\mathcal{T}}^{\{\mu_{1}\mu_{2}\cdots\mu_{M}\}}=\otimes_{n=1}^{M}
U^{\mu_n}_{\{\mu_{1}\mu_{2}\cdots\mu_{M}\}}\Sigma^{\{\mu_{1}\mu_{2}\cdots\mu_{M}\}}\ee
be the high order singular value decomposition of
${\mathcal{T}}^{\{\mu_{1}\mu_{2}\cdots\mu_{M}\}}$. We are readily to
present out main result.\\

{\bf{Theorem:}} For N-qubit quantum systems,
$\Sigma^{\{\mu_{1}\mu_{2}\cdots\mu_{M}\}}$ with
$\{\mu_{1}\mu_{2}\cdots\mu_{M}\}\subset\{1,2,\cdots,N\}$ give a
local unitary classification up to the local symmetry
$\otimes_{n=1}^{M}P^{\mu_n}_{\{\mu_{1}\mu_{2}\cdots\mu_{M}\}}$. For
$2\leq M\leq N$, $P^{\mu_n}_{\{\mu_{1}\mu_{2}\cdots\mu_{M}\}}$ are
formed by a set of orthogonal matrices
$\{V^{\mu_n}_{\{\mu_{1}\mu_{2}\cdots\mu_{M}\}}\}$ and
$U^{\mu_n}_{\{\mu_{1}\mu_{2}\cdots\mu_{M}\}}$ such that
\be\label{pp}P^{\mu_n}_{\{\mu_{1}\mu_{2}\cdots\mu_{M}\}}=(V^{\mu_n}_{\{\mu_{1}\mu_{2}\cdots\mu_{M}\}})^{\dag}
P^{\mu_1}_{\{\mu_{1}\}}U^{\mu_n}_{\{\mu_{1}\mu_{2}\cdots\mu_{M}\}}.\ee\\

{\bf{Proof:}} From the double cover relation between $SU(2)$ and
$SO(3)$, we have that $\rho$ and $\rho'$ are local unitary
equivalent if and only if there exist orthogonal matrices $O^1,
O^2,\cdots,O^N$ such that (\ref{oo}) hold
 for any
$\{\mu_{1}\mu_{2}\cdots\mu_{M}\}\subset\{1,2,\cdots,N\}$.

Denoting
\be\label{sst}{\mathcal{T'}}^{\{\mu_{1}\mu_{2}\cdots\mu_{M}\}}=\otimes_{n=1}^{M}
V^{\mu_n}_{\{\mu_{1}\mu_{2}\cdots\mu_{M}\}}\Lambda^{\{\mu_{1}\mu_{2}\cdots\mu_{M}\}}\ee
the high order singular decompositions of the tensors
${\mathcal{T'}}^{\{\mu_{1}\mu_{2}\cdots\mu_{M}\}}$ and noticing
(\ref{oo}) and (\ref{st}) we have that

\begin{eqnarray*}
\Lambda^{\{\mu_{1}\cdots\mu_{M}\}}=\otimes_{n=1}^{M}(V^{\mu_n}_{\{\mu_{1}\cdots\mu_{M}\}})^{\dag}
O^{\mu_n}U^{\mu_n}_{\{\mu_{1}\cdots\mu_{M}\}}\Sigma^{\{\mu_{1}\cdots\mu_{M}\}}.
\end{eqnarray*}

Define
$$P^{\mu_n}_{\{\mu_{1}\mu_{2}\cdots\mu_{M}\}}=(V^{\mu_n}_{\{\mu_{1}\mu_{2}\cdots\mu_{M}\}})^{\dag}
O^{\mu_n}U^{\mu_n}_{\{\mu_{1}\mu_{2}\cdots\mu_{M}\}}$$ and
$O^{\mu_n}=P_{\{\mu_1\}}^{\mu_n}$. According to the proposition 2 in
\cite{bliu} and the proof therein, we have that
$P^{\mu_n}_{\{\mu_{1}\mu_{2}\cdots\mu_{M}\}}$ must be some local
symmetric operators.

On the other hand, if $\Lambda^{\{\mu_{1}\mu_{2}\cdots\mu_{M}\}}$
and $\Sigma^{\{\mu_{1}\mu_{2}\cdots\mu_{M}\}}$ are related by some
$P^{\mu_n}_{\{\mu_{1}\mu_{2}\cdots\mu_{M}\}}$ with the conditions
listed in the theorem, by selecting the set of orthogonal matrices
$V^{\mu_n}_{\{\mu_{1}\mu_{2}\cdots\mu_{M}\}}$ to be the ones in
the high order singular value decompositions of
${\mathcal{T'}}^{\{\mu_{1}\mu_{2}\cdots\mu_{M}\}}$, one finds that
$$O^{\mu_n}=V^{\mu_n}_{\{\mu_{1}\mu_{2}\cdots\mu_{M}\}}P^{\mu_n}_{\{\mu_{1}\mu_{2}\cdots\mu_{M}\}}
(U^{\mu_n}_{\{\mu_{1}\mu_{2}\cdots\mu_{M}\}})^{\dag}=P_{\{\mu_1\}}^{\mu_n}$$
depends only on $\mu_n$. Therefore we have (\ref{oo}) holds for any
$\{\mu_{1}\mu_{2}\cdots\mu_{M}\}\subset\{1,2,\cdots,N\}$, which ends
the proof. \hfill
\rule{1ex}{1ex}\\

The theorem provides a practical protocol to verify whether two
$N$-qubit mixed states $\rho$ and $\rho'$ are local unitary equivalent or
not:

(1) First write $\rho$ and $\rho'$ in the generalized Bloch
representations (\ref{OS}).

(2) Compute the high order singular decompositions of the tensors
(\ref{st}) and (\ref{sst}). Find all the singular values for
$\{\mu_{1}\mu_{2}\cdots\mu_{M}\}\subset\{1,2,\cdots,N\}$. Since the
tensors we refer here are all with sub dimensional 3, the
computation of the  high order singular value decomposition can be
very simple. If the corresponding singular values are not the same,
$\rho$ and $\rho'$ are not local unitary equivalent.

(3) Otherwise, for any $M\in\{1,2,\cdots,N\}$, check if there exist
some proper $P^{\mu_n}_{\{\mu_{1}\mu_{2}\cdots\mu_{M}\}}$ such that
$$\Sigma^{\{\mu_{1}\mu_{2}\cdots\mu_{M}\}}=\otimes_{n=1}^MP^{\mu_n}_{\{\mu_{1}\mu_{2}\cdots\mu_{M}\}}\Lambda^{\{\mu_{1}\mu_{2}\cdots\mu_{M}\}}$$
and (\ref{oo}) hold, where
$$O^{\mu_n}=V^{\mu_n}_{\{\mu_{1}\mu_{2}\cdots\mu_{M}\}}P^{\mu_n}_{\{\mu_{1}\mu_{2}\cdots\mu_{M}\}}(U^{\mu_n}_{\{\mu_{1}\mu_{2}\cdots\mu_{M}\}})^{\dag}$$
are orthogonal operators depending on the ${\mu_n}$th subsystem
only. If no such $P^{\mu_n}_{\{\mu_{1}\mu_{2}\cdots\mu_{M}\}}$ can
be found, $\rho$ and $\rho'$ are not local unitary equivalent.\\

{\it{Searching the local symmetries for multi-qubit systems:}} From
the above procedures, we see that, since the expression of the core
tensor is only unique up to some local symmetries, the problem
becomes very hard to implement the LU classification. Such problems
exist too in \cite{bliu} for the case of multi-qudit pure states.
Fortunately, by using the iterative property of the conditions
satisfied by the core tensors and taking into account that
${\mathcal{T}}^{\{\mu_{1}\mu_{2}\cdots\mu_{M}\}}$ are real, we are
able to propose a rather simple way to do the step (3) above
operationally. In the following we give a detailed explanation for
searching the local symmetries for multi-qubit mixed states.

We start with the first order tensors
$$\Lambda^{\{\mu_{1}\}}={\mathcal{T}}'^{\{\mu_1\}}=P^{\mu_1}\Sigma^{\{\mu_{1}\}}=P^{\mu_1}{\mathcal{T}}^{\{\mu_1\}}.$$
Any local symmetry $P^{\mu_1}$ transform $\Sigma^{\{\mu_{1}\}}$ to
$\Lambda^{\{\mu_{1}\}}$ can be represented as a rotation. Since the
first order tensors $\Lambda^{\{\mu_{1}\}}$ and
$\Sigma^{\{\mu_{1}\}}$ are 3 dimensional vectors, $P^{\mu_1}$ is a
rotation of $SO(3)$ group with three independent parameters. From
$\Lambda^{\{\mu_{1}\}}=P^{\mu_1}\Sigma^{\{\mu_{1}\}}$, we can reduce
the free parameters in $P^{\mu_1}$ from three to one.
Correspondingly, $O^{\mu_1}=P^{\mu_{1}}$.

From the condition of the second order tensor, one has \be
P^{\mu_1}_{\{\mu_1\mu_2\}}=(V^{\mu_1}_{\{\mu_1\mu_2\}})^{\dag}O^{\mu_1}U^{\mu_1}_{\{\mu_1\mu_2\}}
=(V^{\mu_1}_{\{\mu_1\mu_2\}})^{\dag}P^{\mu_{1}}U^{\mu_1}_{\{\mu_1\mu_2\}}.\nonumber\ee
The free one parameter in $P^{\mu_{1}}$ is further determined by the
equations
\be\Sigma^{\{\mu_1\mu_2\}}=P^{\mu_1}_{\{\mu_1\mu_2\}}\otimes
P^{\mu_2}_{\{\mu_1\mu_2\}}\Lambda^{\{\mu_1\mu_2\}}.\nonumber\ee Then
one checks the third order tensor's condition, the fourth one, ... ,
and all the conditions iteratively. If at least a set of
$\{P^{\mu_n},n=1,2,\cdots,N\}$ satisfies all the conditions in the
above procedures, then we conclude that $\rho$ and $\rho'$ are LU
equivalent. Or we get a negative answer. The corresponding
$U_1,U_2,\cdots, U_N$ can be immediately computed by using the
double cover relation between $SU(2)$ and $SO(3)$ (parameterized as
that in \cite{4396}).

Generally, for any given multi-qubit quantum state $\rho$, we are
able to find the equivalent class of $\rho$ under LU. We first
compute all the tensors
${\mathcal{T}}^{\{\mu_{1}\mu_{2}\cdots\mu_{M}\}}$ and the
corresponding core tensors
$\Sigma^{\{\mu_{1}\mu_{2}\cdots\mu_{M}\}}$, where $1\leq M\leq N$.
Then we parameterize the $3\times3$ orthogonal matrix
$\{P^{\mu_n}_{\{\mu_1\}},n=1,2,\cdots,N\}$ and define (\ref{pp}) for
$2\leq M\leq N$ and for any
$V^{\mu_n}_{\{\mu_{1}\mu_{2}\cdots\mu_{M}\}}$. According to the
theorem, all the
$\{P^{\mu_n}_{\{\mu_{1}\mu_{2}\cdots\mu_{M}\}},n=1,2,\cdots,N\}$ and
the core tensors $\Sigma^{\{\mu_{1}\mu_{2}\cdots\mu_{M}\}}$ form the
class of quantum states that are LU equivalent to $\rho$.

{\it {Example:}} Consider two mixed three-qubit quantum states:
$\rho=\frac{2}{17}(2|\psi_+\ra\la\psi_+|+|001\ra\la001|+|010\ra\la010|+2|011\ra\la011|+\frac{1}{2}|100\ra\la100|+|101\ra\la101|+|110\ra\la110|)$
with $|\psi_+\ra=\frac{1}{\sqrt{2}}(|000\ra+|111\ra)$ and
\begin{eqnarray*}
\sigma=\frac{2}{17}\left(%
    \begin{array}{ccccccccc}
  1& 0& 0& 0& 0& \frac{1}{2}& 0& \frac{1}{2}\\
  0& \frac{3}{2}& 0& \frac{1}{2}& 0& 0& 0& 0\\
  0& 0& 1& 0& 0& -\frac{1}{2}& 0& -\frac{1}{2}\\
  0& \frac{1}{2}& 0& \frac{3}{2}& 0& 0& 0& 0\\
  0& 0& 0& 0& \frac{3}{4}& 0& \frac{1}{4}& 0\\
  \frac{1}{2}& 0& -\frac{1}{2}& 0& 0& 1& 0& 0\\
  0& 0& 0& 0& \frac{1}{4}& 0& \frac{3}{4} & 0\\
  \frac{1}{2}& 0& -\frac{1}{2}& 0& 0& 0& 0 & 1\\
     \end{array}%
    \right).
\end{eqnarray*}

The eigenvalues of $\rho$ and $\sigma$ are both
$\{\frac{2}{17},\frac{2}{17},\frac{2}{17},\frac{2}{17},\frac{4}{17},\frac{4}{17},\frac{1}{17},0\}$,
which means that the two states we considering are degenerate. The
corresponding correlation tensors are listed below:
\begin{eqnarray*}
{\mathcal{T}}^{\{1\}}=(0,0,-\frac{3}{17});\end{eqnarray*}
\begin{eqnarray*}{\mathcal{T}}^{\{2\}}={\mathcal{T}}^{\{3\}}=(0,0,\frac{3}{17});\end{eqnarray*}
\begin{eqnarray*}{\mathcal{T}}^{\{12\}}={\mathcal{T}}^{\{13\}}=(0,0,0,0,0,0,0,0,-\frac{1}{17}),\end{eqnarray*}
\begin{eqnarray*}{\mathcal{T}}^{\{23\}}=(0,0,0,0,0,0,0,0,\frac{1}{17});\end{eqnarray*}
\begin{eqnarray*}{\mathcal{T}}^{\{123\}}&&=(\frac{4}{17},0,0,0,-\frac{4}{17},0,0,0,0,0,
-\frac{4}{17},0,-\frac{4}{17},\\
&&\quad 0,0,0,0,0,0,0,0,0,0,0,0,0,-\frac{3}{17});\end{eqnarray*}
\begin{eqnarray*}{\mathcal{T}}'^{\{1\}}=(0,0,-\frac{3}{17}),
{\mathcal{T}}'^{\{2\}}=(\frac{3}{17},0,0), \end{eqnarray*}
\begin{eqnarray*}{\mathcal{T}}'^{\{3\}}=(0,0,\frac{3}{17});\end{eqnarray*}
\begin{eqnarray*}{\mathcal{T}}'^{\{12\}}=(0,0,0,0,0,0,-\frac{1}{17},0,0),\end{eqnarray*}
\begin{eqnarray*}{\mathcal{T}}'^{\{13\}}=(0,0,0,0,0,0,0,0,-\frac{1}{17}),\end{eqnarray*}
\begin{eqnarray*}{\mathcal{T}}'^{\{23\}}=(0,0,\frac{1}{17},0,0,0,0,0,0);\end{eqnarray*}
\begin{eqnarray*}{\mathcal{T}}'^{\{123\}}&&=(0,0,0,0,-\frac{4}{17},0,-\frac{4}{17},0,0,0,0,0,-\frac{4}{17}
0,\\
&&\quad 0,0,
\frac{4}{17},0,0,0,-\frac{3}{17},0,0,0,0,0,0).\end{eqnarray*} One
can check that all the correlation tensors are just in their core
tensor form. And all the corresponding singular values are the same.
So we come to the step (3) to detect the LU equivalence of $\rho$
and $\sigma$.

To derive $P^1, P^2$ and $P^3$, we parameterize any $2\times 2$
unitary matrix by sketching the representation in \cite{4396} as
\begin{eqnarray*}\label{u}
U=\left(%
    \begin{array}{cc}
  \cos{\theta}e^{i\phi}& \sin{\theta}e^{i\chi}\\
  -\sin{\theta}e^{-i\chi}& \cos{\theta}e^{-i\phi}\\
     \end{array}%
    \right).
\end{eqnarray*}
The corresponding orthogonal matrix derived in (\ref{rr}) can be
written as
\begin{widetext}
\begin{eqnarray*}\label{o}
P=\left(%
    \begin{array}{ccc}
  \cos^2(\theta)\cos(2\phi)-\sin^2(\theta)\cos(2\chi)& \cos^2(\theta)\sin(2\phi)+\sin^2(\theta)\sin(2\chi)& -\sin(2\theta)\cos(\phi+\chi)\\
  \sin^2(\theta)\sin(2\chi)-\cos^2(\theta)\sin(2\phi)& \cos^2(\theta)\cos(2\phi)+\sin^2(\theta)\cos(2\chi)& \sin(2\theta)\sin(\phi+\chi)\\
  \sin(2\theta)\cos(\chi-\phi)& -\sin(2\theta)\sin(\chi-\phi)& \cos(2\theta)\\
     \end{array}%
    \right).
\end{eqnarray*}
\end{widetext}

By solving that ${\mathcal{T}}'^{\{1\}}=P^1{\mathcal{T}}^{\{1\}}$,
${\mathcal{T}}'^{\{2\}}=P^2{\mathcal{T}}^{\{2\}}$, and
${\mathcal{T}}'^{\{3\}}=P^3{\mathcal{T}}^{\{3\}}$, one obtains the
corresponding parameters to be $\{\theta_1=0\},
\{\theta_2=\frac{\pi}{4},\phi_2=\pi-\chi_2\}$ and $\{\theta_3=0\}$.

Then we simplify $P^1, P^2$ and $P^3$ to be

\begin{flalign}
&P^1=\left(%
    \begin{array}{ccc}
  \cos(2\phi_1)& \sin(2\phi_1)& 0\\
  -\sin(2\phi_1)& \cos(2\phi_1)& 0\\
  0& 0& 1\\
     \end{array}%
    \right),\nonumber\\[3mm]
&P^2=\left(%
    \begin{array}{ccc}
  0& 0& 1\\
  \sin(2\chi)&  \cos(2\chi)& 0\\
  -\cos(2\chi)& \sin(2\chi)& 0\\
     \end{array}%
    \right),\nonumber\\[3mm]
&P^3=\left(%
    \begin{array}{ccc}
  \cos(2\phi_3)& \sin(2\phi_3)& 0\\
  -\sin(2\phi_3)& \cos(2\phi_3)& 0\\
  0& 0& 1\\
     \end{array}%
    \right).\nonumber
\end{flalign}

By the rest conditions that the second and third order core tensors
should satisfy, i.e.
\begin{eqnarray*}&&{\mathcal{T}}'^{\{12\}}=P^1\otimes P^2
{\mathcal{T}}^{\{12\}};\\
&& {\mathcal{T}}'^{\{13\}}=P^1\otimes P^3 {\mathcal{T}}^{\{13\}};\\
&& {\mathcal{T}}'^{\{23\}}=P^2\otimes P^3
{\mathcal{T}}^{\{23\}}\end{eqnarray*} and
$${\mathcal{T}}'^{\{123\}}=P^1\otimes P^2\otimes P^3
{\mathcal{T}}^{\{123\}},$$ one can find the final solutions for
$P^1, P^2$ and $P^3$ with parameters satisfying
$\phi_1-\chi_2+\phi_3=0$.

Finally we obtain that the survival local symmetries $P^1, P^2$ and
$P^3$ must be with the conditions $\{\theta_1=0\},
\{\theta_2=\frac{\pi}{4},\phi_2=\pi-\chi_2\}, \{\theta_3=0\}$ and
$\phi_1-\chi_2+\phi_3=0$. For example, one can select
$\theta_1=\phi_1=0, \theta_2=\frac{\pi}{4}, \phi_2=\pi, \chi_2=0$
and $\theta_3=\phi_3=0$. Then we have that $P^1=P^3=I$ and $P^2$ be
an anti-diagonal matrix with entries $\{1,1,-1\}$. The corresponding
local unitary operators that transform $\rho$ to $\sigma$ are then
computed to be  $U_1=U_3=I$ and $U_2=\{\{\frac{\sqrt{2}}{2}e^{i\pi},
\frac{\sqrt{2}}{2}\},
  \{-\frac{\sqrt{2}}{2}, \frac{\sqrt{2}}{2}e^{-i\pi}\}\}$.\\

{\it Conclusions and Remarks:} It is a basic and fundamental
question to classify quantum states under local unitary operations.
The problem has been figured out in \cite{mqubit,bliu} for pure
multipartite quantum states. However, it is much more difficult to
classify mixed quantum states under LU transformations than that for
pure states. Operational methods have been presented only for
non-degenerated bipartite states. Although the authors in \cite{jpa}
have shown that the problem of mixed states can be reduced to the
one of pure states in terms of the purification of mixed states
mathematically, the protocol is rather not operational. In this
letter we have provided an operational way to verify and classify
quantum states by using the generalized Bloch representation in
terms of the generators of $SU(d)$. Since the tensors in the
representation can be all determined by some local quantum
mechanical observables, the method is experimentally feasible. Our
criterion is both sufficient and necessary for multi-qubit quantum
systems, thus gives rise to a complete classification of multi-qubit
quantum states under LU transformations.

\smallskip
\noindent{\bf Acknowledgments}\, \, This work is supported by the
NSFC 11105226, 11275131; the CSC; the Fundamental Research Funds for
the Central Universities No.12CX04079A, No.24720122013; Research
Award Fund for outstanding young scientists of Shandong Province
No.BS2012DX045.


\begin{thebibliography}{99}
\bibitem{nielsen}M.A. Nielsen and I.L. Chuang, Quantum Computation and Quantum
Information. Cambridge: Cambridge University Press, (2000).

\bibitem{1} D. Gottesman, Ph.D. thesis, Caltech [arXiv:quant-ph/ 9705052].


\bibitem{2} R. Raussendorf and H. J. Briegel, Phys. Rev. Lett. 86, 5188
(2001).

\bibitem{3} L. Amico et al., Rev. Mod. Phys. 80,
517 (2008); F. Verstraete et al., Adv. Phys. 57, 143 (2008), and
references therein.

\bibitem{4} R. Horodecki et al., Rev. Mod. Phys. 81, 865 (2009).

\bibitem{eof1} C. H. Bennett, D. P. DiVincenzo, J. A. Smolin, and W. K.
Wootters, Phys. Rev. A 54, 3824 (1996); C. H. Bennett, H. J.
Bernstein, S. Popescu, and B. Schumacher, ibid. 53, 2046 (1996); V.
Vedral, M. B. Plenio, M. A. Rippin, and P. L. Knight, Phys. Rev.
Lett. 78, 2275 (1997).

\bibitem{eof2} W. K. Wootters, Phys. Rev. Lett. 80, 2245 (1998).

\bibitem{bell1} R. Horodecki, P. Horodecki, and M. Horodecki, Phys. Lett. A 200,
340 (1995).

\bibitem{bell2} R. F. Werner and M. M. Wolf, Phys. Rev. A 64, 032112 (2001).

\bibitem{bell3} M. Zukowski and C. Brukner, Phys. Rev. Lett. 88, 210401
(2002).

\bibitem{bell4} M. Li and S.M. Fei, Phys. Rev. A 86, 052119 (2012).


\bibitem{tel1} M. Horodecki, P. Horodecki, and R. Horodecki,
Phys. Rev. A, 60, 1888(1999).

\bibitem{tel2} S. Albeverio, S. M. Fei, and W. L. Yang, Phys. Rev. A, 66,
012301(2002).

\bibitem{makhlin} Y. Makhlin, Quant. Info. Proc. 1, 243 (2002).

\bibitem{linden} N. Linden, S. Popescu and A. Sudbery, Phys. Rev. Lett. 83, 243 (1999).

\bibitem{Linden} N. Linden and S. Popescu, Fortsch. Phys. 46, 567 (1998).

\bibitem{SFY} S. Albeverio, S.M. Fei, P.Parashar,W.L.Yang, Phys. Rev. A 68, 010303 (2003).

\bibitem{SFG} S. Albeverio, S.M. Fei, and D.Goswami, Phys. Lett. A 340, 37 (2005).

\bibitem{SFW} B.Z. Sun, S.M. Fei, X.Q. Li-Jost and Z.X.Wang, J. Phys. A 39, L43-L47(2006).

\bibitem{SCFW} S. Albeverio, L. Cattaneo, S.M. Fei and X.H. Wang, Int. J. Quant. Inform. 3, 603 (2005).

\bibitem{mqubit} B. Kraus, Phys. Rev. Lett. 104, 020504(2010); Phys. Rev. A 82, 032121 (2010).

\bibitem{bliu} B. Liu, J.L. Li, X. Li, C.F. Qiao, Phys. Rev. Lett.108, 050501 (2012).

\bibitem{zhou} C. Zhou, T.G. Zhang, S.M. Fei, N. Jing, and X. Li-Jost, Phys. Rev. A 86(R), 010303 (2012).

\bibitem{zhang} T.G. Zhang, M.J. Zhao, M. Li, S.M. Fei, and X. Li-Jost, Phys. Rev. A, 88, 042304 (2013)

\bibitem{4396} J. Schlienz and G. Mahler, Phys. Rev. A 52, 4396 (1995).

\bibitem{hassan} A. Saif M. Hassan and Pramod S. Joag, Quant. Info. Comput. 8, 0773(2008).

\bibitem{siam} L. D. Lathauwer, B.D. Moor, and J. Vandewalle, SIAM
J. Matrix Anal. Appl.21, 1253(2000);\\
Tamara G. Kolda and Brett W. Bader, SIAM Rev.51, 455(2009).


\bibitem{fenlei} W. Dur, G. Vidal, and J.I. Cirac, Phys. Rev. A 62, 062314
(2000).

\bibitem{jpa} J.L. Li and C.F. Qiao, J. Phys. A: Math. Theor. 46, 075301
(2013).

\end{thebibliography}
\end{document}